\newcommand{\Gd}{Gd$_{2}$Ti$_{2}$O$_{7}$}

\newcommand{\e}{{\rm e}}

\documentclass[twocolumn,showpacs,preprintnumbers,amsmath,amssymb,prl]{revtex4}

\usepackage{graphicx}
\usepackage{dcolumn}
\usepackage{bm}

\begin{document}

\preprint{}

\title{Theory of Two-Step Magnetic Ordering Phenomena in a Geometrically
Frustrated Heisenberg Pyrochlore Antiferromagnet with 
Long Range Dipolar Interactions}

\author{Matthew Enjalran$^1$}
 \email{enjalran@gandalf.uwaterloo.ca}
\author{Michel J.P. Gingras$^{1,2}$}
 \email{gingras@gandalf.uwaterloo.ca}
\affiliation{$^1$Department of Physics, University of Waterloo, 
Ontario, N2L 3G1, Canada \\ 
$^2$Canadian Institute for Advanced Research, 180 Dundas Street West,
Toronto, Ontario, M5G 1Z8,Canada}

\date{\today}

\begin{abstract}
A model for a Heisenberg antiferromagnet on a pyrochlore 
lattice with exchange and dipole-dipole interactions is studied 
via mean-field theory. 
In treating the dipoles by use of the 
Ewald method, a soft (critical) mode with a unique
ordering wave vector is selected for 
all strengths of the dipole-dipole coupling. For weak dipoles a 
partially ordered, three sublattice spin structure (P state), 
with ${\bm q}_{\rm ord}= \frac{1}{2}  \frac{1}{2} \frac{1}{2}$, 
is selected. A fully ordered, four sublattice spin structure (F state), 
with ${\bm q}= 0 0 0$, competes with the P state and 
becomes the stable structure as the temperature 
is reduced. Our results are compared against other
theoretical calculations and connection to recent experimental
results for \Gd\ are discussed.
\end{abstract}

\maketitle 


Frustration is ubiquitous in condensed matter and arises whenever a system
cannot minimize its total classical ground state energy by 
minimizing the energy
of its pairwise interactions, pair by pair. 
In the context of magnetism, 
frustration commonly occurs in triangular or tetrahedral unit 
based lattice structures with antiferromagnetic (AFM) interactions, 
geometric frustration, and 
prevents the magnetic moments (spins) from pointing 
antiparallel to each other. Two and three
dimensional lattices of corner sharing triangles or tetrahedra
with nearest neighbor AFM Heisenberg exchange are highly frustrated and 
are particularly interesting. 
There, mean-field theory (MFT) finds a massive degeneracy of soft (critical) modes
with no unique preferred state chosen at some critical 
temperature \cite{rbs-mft,gingras_cjp}.
As a consequence, frustrated systems exhibit a 
propensity for thermal and quantum 
fluctuations \cite{qaf-moessner,qaf-hermele}.
The possibility that exotic novel quantum states arise in 
frustrated quantum spin systems is currently attracting much attention.

\begin{figure}[!ht]
\begin{center}
\includegraphics[width=2.0in,height=1.8in]{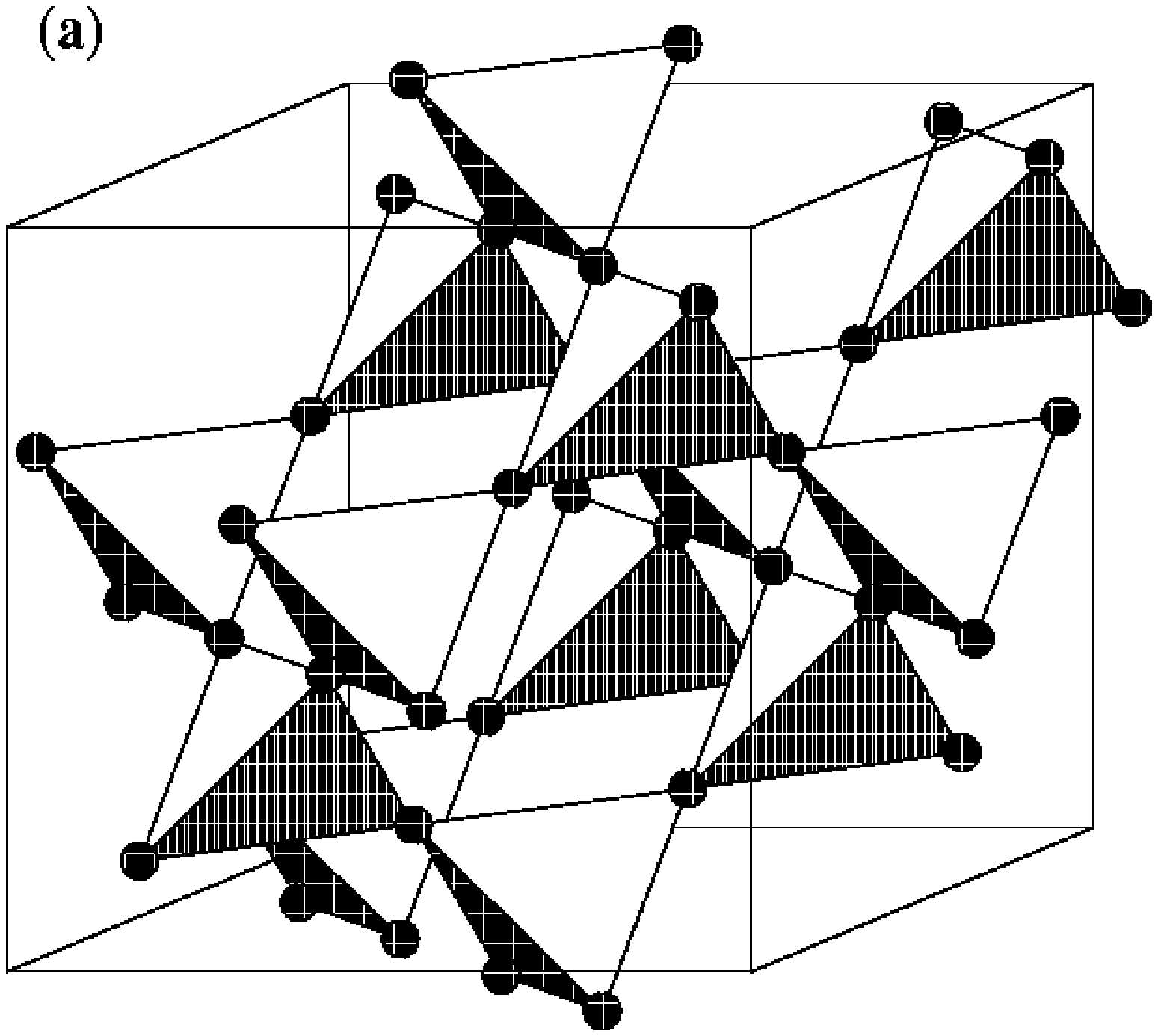}
\includegraphics[width=2.5in,height=1.2in]{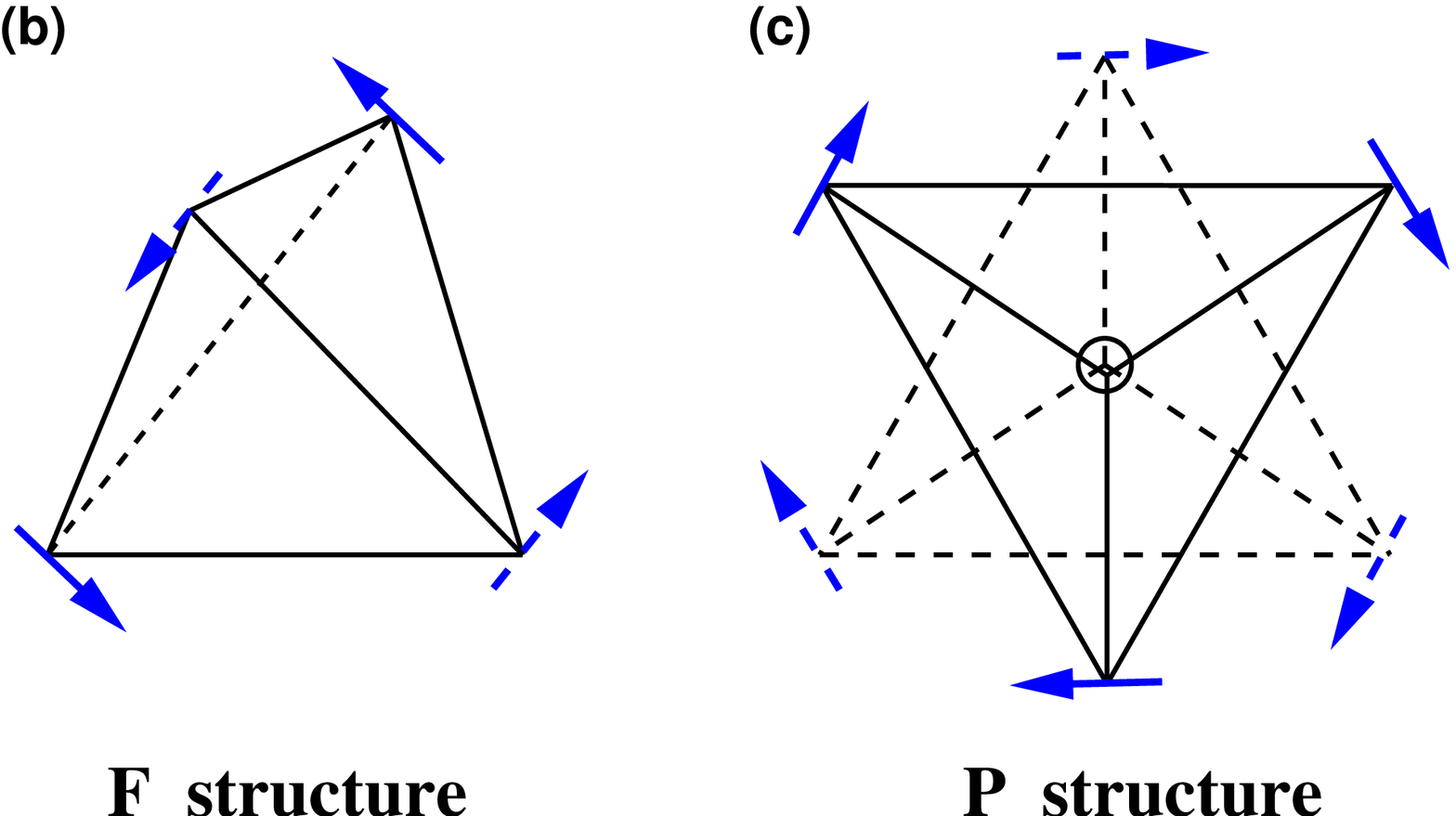}
\caption{(a) The pyrochlore lattice as a fcc Bravais 
lattice with a four atom basis. 
(b) The fully ordered ${\bm q} = 000$ equal moment spin structure 
(F state) about a single tetrahedron. Each spin is parallel 
to an opposite side of a tetrahedron, 
each tetrahedron has zero net moment, $\sum_{a=1}^{4} {\bm S}^a = 0$. 
(c) The partially ordered
${\bm q} = \frac{1}{2} \frac{1}{2} \frac{1}{2}$ spin structure (P state) 
projected onto the plane perpendicular
to the $[111]$ direction. 
The solid (dashed) lines represent a tetrahedron 
projected out of (in to) the plane of the paper. 
The three sublattices with equal moments lie
in a plane (kagom\'{e}) and sum to zero with the fourth interstitial site, 
encircled, paramagnetic. There is a reversal of spin direction between spins 
in adjacent layers.}
\label{fig-pyro}
\end{center}
\end{figure}

It is generally expected that perturbations to the nearest 
neighbor spin Hamiltonian select a unique ordered
state. In the insulating pyrochlore systems R$_2$M$_2$O$_7$, 
where R$^{3+}$ is a magnetic trivalent
rare earth ion (R=Gd, Tb, Ho, Dy), and M=Ti$^{4+}$ or Sn$^{4+}$,
non-magnetic (Fig.~\ref{fig-pyro}a), 
the leading perturbation to the nearest-neighbor Hamiltonian is the long-range
magnetic dipole-dipole interaction. Due to its (i) long-range $1/r^3$ and
(ii) anisotropic ${\bm r}\cdot {\bm S}$ spin$-$space coupling nature, 
one would naively
expect dipolar interactions to select a unique and robust soft 
magnetization density (mode) at an ordering wave vector ${\bm q}_{\rm ord}$.
However, explicit calculations in two specific 
cases have found dramatically different behavior.

In the Ho$_2$(Ti,Sn)$_2$O$_7$ and Dy$_2$Ti$_2$O$_7$ pyrochlores with local
$\langle 111 \rangle$ Ising spins, it has been found that 
the $1/r^3$ dipolar interactions are to a very
large extend self-screened \cite{gingras_cjp}. 
As a result, the system does not develop long-range 
order at a critical temperature 
$T_c \sim D_{\rm nn}$,
where $D_{\rm nn} \sim 2$ K is the dipolar energy scale at 
nearest-neighbors. Instead, 
the system develops an extensive quasi-degenerate ``spin ice'' 
state at $T \sim D_{\rm nn}$ 
analogous to the proton disordered state in common ice water with 
the same residual extensive Pauling entropy \cite{ramirez-nature,rev-SI}. 
Interestingly, detailed calculations 
(mean-field \cite{gingras_cjp} and Monte Carlo \cite{SI-lro})
find that the dipolar interactions do	
not produce an exact symmetry and thus allows for the weak 
selection of an ordered state at ordering wave vector ${\bm q}_{ord} = 001$.
Monte Carlo simulations employing nonlocal loop dynamics 
find a first order transition, removing all residual entropy, to
the ${\bm q}_{ord} = 001$ long-range ordered state at 
the critical temperature $T_c/D_{\rm nn} \approx 0.07 \ll 1$ \cite{SI-lro}.
 
In the Gd$_2$(Ti,Sn)$_2$O$_7$ pyrochlores, the Gd$^{3+}$ 
ions have $S=7/2$ and $L=0$, hence negligible single-ion
anisotropy. They are, therefore, good realizations of 
antiferromagnetically coupled Heisenberg spins with 
weak long-range magnetic dipole-dipole interactions
($20\%$ of nearest-neighbor AFM exchange) \cite{gdtio-raju}.
Here too, the effect of dipolar interactions are far from obvious.
Two independent mean-field (Gaussian approximation)
calculations have found that long-range dipole-dipole interactions
truncated at large distance (i.e., 100 nearest-neighbors) give rise to
a line of degenerate soft modes along the cubic $\langle 111 \rangle$ 
direction \cite{gdtio-raju,gdtio-palmerchalker}.
Palmer and Chalker argued that including quartic terms in the theory leads
to the selection of a four sublattice N\'eel ordered state 
at ${\bm q}_{ord} = 000$, see Fig.~\ref{fig-pyro}b.
Experiments provide a different picture: 
(i) Neutron diffraction measurements  
at $T=50$mK find a partially ordered phase at
${\bm q}=\frac{1}{2} \frac{1}{2} \frac{1}{2}$, see Fig.~\ref{fig-pyro}c, 
with one disordered sublattice \cite{gdtio-bramwell}.
(ii) M\"{o}ssbauer experiments \cite{gdtio-mossbauer} 
find strong evidence 
for a paramagnetic (PM) fluctuating
spin down to $35$mK, too.
(iii) Specific heat measurements \cite{gdtio-ramirez} find two transitions at
$T=0.7$K and $T=1.0$K.  All these experimental results are seemingly 
incompatible with the prediction of Ref.~\onlinecite{gdtio-palmerchalker}.
In view of these perplexing results from theory and experiments, 
we have revisited the mean-field calculations
of Refs. \onlinecite{gdtio-raju} and \onlinecite{gdtio-palmerchalker}. 
We have found that the degeneracy of soft modes
along the ${\bm q}= 111$ direction reported in \onlinecite{gdtio-raju} and 
\onlinecite{gdtio-palmerchalker}
is only approximate.
Instead, we find that long-range dipolar interactions, 
when properly re-summed to infinite distance, 
give rise to a weakly selected unique soft mode
at ${\bm q}_{ord}= \frac{1}{2} \frac{1}{2} \frac{1}{2}$ with one disordered
sublattice, similar to what is found in experiments 
(Fig.~\ref{fig-pyro}c) \cite{gdtio-bramwell}. 
Just $0.7$\% below that first transition, we find a second 
transition to the ${\bm q}= 000$ state of Fig.~\ref{fig-pyro}b. 
We also find that weak second and third nearest neighbor exchange 
only acts to suppress or enhance the transition temperature 
of the ${\bm q}= 000$ mode relative to that of the  
${\bm q}_{ord}= \frac{1}{2} \frac{1}{2} \frac{1}{2}$ state.       


We consider a model for large ${\bm S}$-spins on a pyrochlore lattice 
with exchange and dipolar interactions but negligible single-ion 
anisotropy. 
The appropriate Heisenberg Hamiltonian, $H$, is given by
$H = -\frac{1}{2}\sum_{i,j} \sum_{a,b} \sum_{u,v}
{\mathcal J}^{ab}_{uv}(i,j) S_{i}^{a,u} S_{j}^{b,v}$,
where indices $(i,j)$ denote fcc Bravais lattice points, $(a,b)$ 
describe the tetrahedral basis, and $(u,v)$ define the components 
of classical Heisenberg spins, $S_{i}^{a,u}$. The 
${\mathcal J}^{ab}_{uv}(i,j)$ interaction 
matrix contains exchange and dipolar contributions and is given by  
${\mathcal J}^{ab}_{uv}(i,j)  =  
J_N \delta_{ij, N{\rm nn} }\delta^{uv} - D_{\rm dd}  
\left[ {\delta^{uv}}{|{\bm R}_{ij}^{ab}|^{-3}}
- {3 {\bm R}_{ij}^{ab,u} {\bm R}_{ij}^{ab,v}}{|{\bm R}_{ij}^{ab}|^{-5}} \right]$, 
where $J_N \delta_{ij, N{\rm nn} }$ restricts the exchange to the $N$-th 
nearest neighbor distance, $D_{\rm dd}\equiv DR_{\rm nn}^3$ with  
$D=\mu_{\rm o}\mu^2 / 4\pi R_{\rm nn}^3$,
$\mu$ is the magnetic moment occupying the pyrochlore lattice site, 
$R_{\rm nn}$ is the nearest neighbor distance, and 
${\bm R}_{ij}^{ab}$ is the vector separation 
between spins ${\bm S}_{i}^{a}$ and ${\bm S}_{j}^{b}$. 
The nearest neighbor exchange, $J_1$, sets 
the energy scale ($J_1<0$ is AFM). 
The effects of weak long-range exchange ($J_2$ and $J_3$) 
are also considered.
  

We first investigate the soft-mode spectrum of $H$
by studying the momentum dependence of the static susceptibility,
$\chi({\bm q})$ within MFT \cite{mft-neutron,SI-melko}. 
The mean-field free energy is 
${\mathcal F}={\rm Tr}\{\rho H \} + T{\rm Tr}\{\rho \ln \rho \}$ 
\cite{rbs-mft,gingras_cjp}, 
where ${\rm Tr}$ is a trace over spin configurations and $\rho$
is the mean-field (one-particle) density matrix. The expression 
for ${\mathcal F}$ is expanded to quadratic order 
in ${\bm m}_i^a$, the vector order parameter,  
Fourier transformed, and then diagonalized 
by a normal mode transformation. 
The diagonal form of ${\mathcal F}$ is used to calculate the correlation 
function $\langle {\bm S}_i^a \cdot {\bm S}_j^b \rangle$ in the
definition of $\chi({\bm q})$,  
\begin{equation}
\label{eq-chiq}
\chi({\bm q}) = (\beta/N_{\rm cell}) \sum_{(i,a),(j,b)} 
\langle {\bm S}_i^a \cdot {\bm S}_j^b \rangle 
\e^{\imath {\bm q} \cdot {\bm R}_{ij}^{ab}} ,
\end{equation}
where $\beta = 1/T$ in units of $k_{\rm B}=1$ and $N_{\rm cell}$ is
the number of Bravais lattice points, and ${\bm q}$ is wave vector in 
the first Brillouin zone. 
The final form for the ${\bm q}$-dependent susceptibility is 
$\chi({\bm q})/N_{\rm cell} = (\beta/3)
\sum_{\alpha,\mu} |{\bm F}^{\alpha}_{\mu}({\bm q})|^2 
(1-\lambda^{\alpha}_{\mu}({\bm q})/3T)^{-1}$,
where the Greek indices label the normal modes ($\alpha=1,2,3,4$ 
and $\mu=1,2,3$) at each value of ${\bm q}$, and 
${\bm F}^{\alpha}_{\mu}({\bm q})
=\sum_{a,u}U_{u,\mu}^{a,\alpha}({\bm q})
=\sum_{a}{\bm U}_{\mu}^{a,\alpha}({\bm q})$ 
is a 3-component vector. 
$U({\bm q})$ is the unitary 
matrix that diagonalizes ${\mathcal J}({\bm q})$ in the sublattice space 
with eigenvalues $\lambda^{\alpha}_{\mu}({\bm q})$. The static 
susceptibility is a maximum at the ordering wave vector, 
${\bm q}_{\rm ord}$, which also defines the maximum eigenvalue, 
i.e., $\lambda^{\rm max}({\bm q}_{\rm ord})
={\rm max}_{\bm q}\{\lambda^{\rm max}({\bm q})\}$, where 
$\lambda^{\rm max}({\bm q})$ is the maximum eigenvalue at 
${\bm q}$ and ${\rm max}_{\bm q}$ selects the global maximum for all
${\bm q}$. The mean-field critical temperature is given by, 
$T_c^{\rm MF} = \lambda^{\rm max}({\bm q}_{\rm ord})/3$.   

\begin{figure}[!ht]
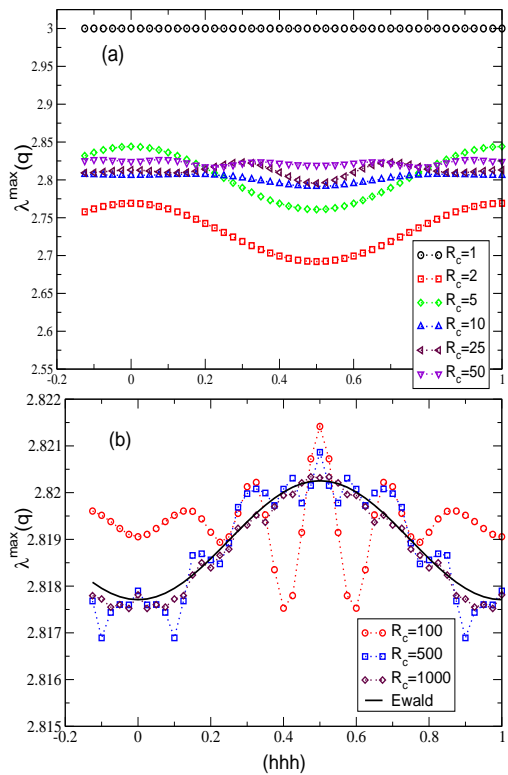

\begin{center}
\includegraphics[width=2.6in,height=2.0in]{fig2a.eps}
\includegraphics[width=2.6in,height=2.0in]{fig2b.eps}
\caption{Soft-mode spectrum along $(hhh)$ as a function of dipolar
cut off, $R_c$, with $D/|J_1|=0.20$. (a) The spectrum develops
$R_c$ dependent local maxima for $R_c > 1$. (b) The fluctuations in 
$\lambda^{\rm max}(hhh)$ are smoothed out as $R_c$ increases with
the $R_c \rightarrow \infty$ limit obtained with the Ewald method. 
In the Ewald limit, the critical mode falls at 
${\bm q}_{\rm ord}= \frac{1}{2}  \frac{1}{2} \frac{1}{2}$.} 
\label{fig-softmode}
\end{center}
\end{figure}

To find ${\bm q}_{\rm ord}$ for our model, we search for the maximum in
the soft mode spectrum ($\lambda^{\rm max}({\bm q})$). 
We consider AFM first 
neighbor exchange, $J_1=-1.0$, 
and a variable dipole-dipole interaction strength, $D$. 
The dipolar interactions in ${\mathcal J}({\bm q})$
are summed directly to a cut off separation distance, 
$R_c$, measured in units of
the nearest neighbor distance and to the infinite range limit 
through the use of the Ewald method \cite{mft-neutron}.
The regime of weak dipole-dipole interactions is of
greatest interest to us because the value $D/|J_1|=0.20$ 
faithfully approximates \Gd \cite{gdtio-raju}.
At this point, we find a spectrum in the $(hhl)$ plane that 
\textit{appears} degenerate along the $(hhh)$ diagonal, in 
agreement with earlier works \cite{gdtio-raju,gdtio-palmerchalker}
(Fig. 6 in Ref.~\onlinecite{gdtio-raju}).
However, when we examine the spectrum along the $(hhh)$ diagonal
in greater detail, we observe the selection of a 
global maximum as one increases the cut off distance of the dipolar
lattice sum. Figure \ref{fig-softmode} displays results for 
several values of $R_c$ and the Ewald limit. An ordering 
wave vector occurs at 
${\bm q}_{\rm ord}= \frac{1}{2}  \frac{1}{2} \frac{1}{2}$ 
(in units of $2\pi/a$), 
and emerges from a smoothly varying spectrum along $(hhh)$ in the
Ewald limit. We note that for dipole-dipole interactions summed
to $R_c < 500$, $\lambda^{\rm max}({\bm q})$ displays 
substantial variations making it difficult to ascertain the 
existence of a critical ordering wave vector. These fluctuations are 
damped with increasing $R_c$ and approach the Ewald limit for
$R_c > 1000$. A similar sensitivity of the soft-mode spectrum
to the dipole-dipole cut off has been observed in the 
spin ice materials\cite{mft-neutron}. 
In a broader expanse of phase space, we find three regions 
for an ordering wave vector: 
(i) ${\bm q}_{\rm ord}=\frac{1}{2}  \frac{1}{2} \frac{1}{2}$ 
for $D/|J_1| \leq 5.60$, 
(ii) ${\bm q}_{\rm ord}= 001$ for $5.60 < D/|J_1| < 5.85$, 
(iii) ${\bm q}_{\rm ord}= 000$ for $D/|J_1| \geq 5.85$.
Earlier work that imposed a cut off on the dipole-dipole lattice
sum found a degenerate spectrum along the $(hhh)$ diagonal for 
$D/|J_1| \leq 5.70$, with the selection of an 
ordering wave vector \textit{near} 
$(000)$ for larger values of $D$, (see Fig. 2 in 
Ref.~\onlinecite{gdtio-palmerchalker}).

The spin structure for ${\bm q}_{\rm ord}=\frac{1}{2} \frac{1}{2} \frac{1}{2}$
can be obtained from the corresponding eigenvectors, 
$U_{u,\mu}^{a,\alpha}(\frac{1}{2}  \frac{1}{2} \frac{1}{2})$. About a single
tetrahedron one infers that three sublattices have 
equal moments with all spins confined to a plane and a fourth
sublattice that is PM, i.e., zero moment,
see Fig.~\ref{fig-pyro}c. 
When this configuration is extended over the pyrochlore lattice and
modulated by ${\bm q}_{\rm ord}$, one has
ordered kagom\'{e} planes along the $[111]$ direction with a 
reversal of spin direction in alternating planes, all  
interstitial sites are PM. This 
partially ordered three sublattice structure 
(P state) has been observed in neutron scattering 
experiments\cite{gdtio-bramwell} and is 
supported by M\"{o}ssbauer\cite{gdtio-mossbauer} 
experiments on \Gd.
The mean-field transition temperature 
to this state is 
$T_c^{\rm P}=\lambda^{\rm max}(\frac{1}{2} \frac{1}{2} \frac{1}{2})/3
=0.94008|J_1|$
($D/|J_1|=0.2$).

The mode ${\bm q}= 000$ corresponds to a 
fully ordered (four sublattice) equal moment structure 
(F state) initially predicted to be the ground state of 
a $D/|J_1| = 0.20$ model \cite{gdtio-palmerchalker}. 
We find the limit of stability for the paramagnetic state
against an ordering to the F state occurs at  
$T_c^{\rm F}=\lambda^{\rm max}(000)/3=0.93923|J_1|$, 
which is very close to $T_c^{\rm P}=0.94008|J_1|$ \cite{numerics}. 
Given the close proximity of 
$T_c^{\rm P}$ and $T_c^{\rm F}$ at the Gaussian level,
we study the evolution of the respective   
free energies as a function of temperature below both values of $T_c$.


The real-space mean-field theory (RSMFT)   
begins with the mean-field decomposition 
$(S_i^{a,u}-m_i^{a,u})(S_j^{b,v}-m_j^{b,v})=0$ applied 
to $H$. The free energy is determined from 
$f=-(1/\beta) \ln Z_{\rm MF}$, where $Z_{\rm MF}$ is the 
RSMFT partition function. For our model, one obtains
\begin{equation}
\label{eq-f}
f = \frac{1}{2} \sum_{i,j}\sum_{a,b}\sum_{u,v} {\mathcal J}^{ab}_{uv}(i,j)
m_i^{a,u}m_j^{b,v} - \frac{1}{\beta} \sum_{i,a} \ln Z_i^a \; ,
\end{equation}
where $Z_i^a=(4\pi/\beta|{\bm B}_i^a|) \sinh(\beta|{\bm B}_i^a|)$
is the single-site partition function and ${\bm B}_i^a$ is the local
field. The sublattice magnetization ${\bm m}_i^a$ (order parameter) is obtained
from differentiation of $f$ with respect to local 
${\bf B}_i^a$ field. One finds
${\bm m}_i^a = \frac{{\bm B}_i^a}{|{\bm B}_i^a|}
\left[\coth(\beta |{\bm B}_i^a|) - \frac{1}{\beta |{\bm B}_i^a|} \right]$, 
where
the field at site $(i,a)$ is due to all the other moments in the system 
acting at site $(i,a)$, i.e.,   
${\bm B}_i^a=\sum_u B_i^{a,u} = \sum_u 
\sum_{j,b,v} {\mathcal J}^{ab}_{uv}(i,j) m_j^{b,v}$.
It is convenient to factor out the magnitude of the
order parameter and write, ${\bm B}_i^a=m{\bm b}_i^a$, where
${\bm m}_i^a=m{\hat n}_i^a$ and ${\hat n}_i^a$ is a unit vector
that defines the orientation of the order parameter at site
$(i,a)$. Using this form for ${\bm B}_i^a$, one
expands the right-hand-side of the above expression for 
${\bm m}_i^a$ for small $m$ to obtain $T_c$ for RSMFT, 
$T_c^{\rm MF} = |{\bm b}_i^a|/3$. 

\begin{figure}[!ht]
\begin{center}
\includegraphics[width=3.0in,height=2.4in]{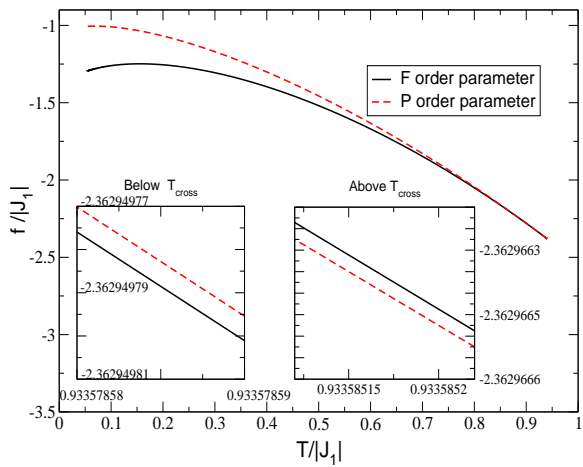}
\caption{Free energy of the P and F order parameters for 
a $L=4$ lattice with $D/|J_1|=0.20$. At 
high temperatures the free energy curves appear to lie on 
top of one another with the F state clearly preferred 
at lower temperatures. 
The insets show the free energy above (P mode favored) 
and below (F mode favored) a crossing temperature 
$T_{\rm cross}/|J_1|\approx 0.93358$.}
\label{fig-freeE}
\end{center}
\end{figure}

We work on a lattice of $L\times L\times L$ cubic cells, or 
$16L^3$ spins, with periodic boundary conditions and 
$D/|J_1|=0.20$. 
The lattice is tiled with one of the above order parameters.  
We consider systems of size $L=4,6$ and calculate the 
dipole-dipole interactions via the Ewald method in real-space \cite{SI-melko}.
Our procedure is to solve self-consistently for the sublattice magnetization, 
${\bm m}_i^a$, at temperature $T/|J_1|$ and 
then calculate the corresponding free energy from Eq.~\ref{eq-f},
i.e., $f/|J_1|$. 
The real-space transition temperatures are 
$T_c^{\rm P}=|{\bm b}_i^{a({\rm P})}|/3=0.94002|J_1|$
and $T_c^{\rm F}=|{\bm b}_i^{a({\rm F})}|/3=0.93917|J_1|$ 
for $L=6$, in good agreement with the above 
${\bm q}$-space results. 
For $L=8$, the real-space and ${\bm q}$-space results for
$T_c^{\rm P}$ and $T_c^{\rm F}$ agree within $10^{-7}$.
Free energy versus temperature curves for the F and P  
ordered states on a $L=4$ lattice are displayed in 
Fig.~\ref{fig-freeE}. 
At high temperatures, near either $T_c^{\rm P}$ or $T_c^{\rm F}$, 
the free energy curves appear degenerate. 
However, the insets to Fig.~\ref{fig-freeE} demonstrate that the 
respective curves actually cross. We performed linear 
fits to points from the two curves about the crossing temperature.
Solving these equations we find the temperature 
$T_{\rm cross} \approx 0.93358|J_1|$. Above $T_{\rm cross}$, 
the P state has a lower free energy and is thus the preferred state. 
Below $T_{\rm cross}$, the free energy of the F state drops 
below that of the P state
and eventually becomes the stable ground state of the system. 
We find $T=0$ energies per spin of  
$E_o^{\rm F}=-1.40856|J_1|$ and $E_o^{\rm P}=-1.05737|J_1|$. 
Hence, in this very simple mean-field 
picture, the three sublattice P state is the initially preferred 
structure at high temperatures, 
but the nearby four sublattice F state competes 
with it and very rapidly wins out as the temperature is lowered.  

The proximity of the P and F states in MFT 
suggests it would be difficult to realize their 
experimental differentiation (or in Monte Carlo simulations).
However, the inclusion of weak \cite{exptJ-greedan} 
long-range exchange in our model separates the 
transition temperatures of these two states and effectively 
reduces their competition.
We note that a $J_2$ or $J_3$ has no effect on the 
transition temperature of the P state \cite{gdtio-shastry}. 
In the Fourier transform of the $J_2$ and $J_3$ matrices, 
the particular combination of the lattice symmetry and the 
$q= \pi\pi\pi$ phase yields zero matrix elements 
for $J_2(\pi\pi\pi)$ and $J_3(\pi\pi\pi)$ in 
the above calculation of $\chi({\bm  q})$ at ${\bm q}= \pi\pi\pi$. 
Hence, we find that $J_3=0.02J_1$ (AFM) suppresses the F 
state to $T_c^{\rm F}=0.859237|J_1|$ ($L=4$). 
A small FM $J_2$ has a similar effect but of reduced magnitude. 
Conversely, a small AFM $J_2$ or 
FM $J_3$ enhances the F state by raising 
its transition temperature above $T_c^{\rm P}$. 

Hence, experimentally reasonable values
for long-range exchange \cite{exptJ-greedan} 
could allow for the detection of
two well-separated transitions in \Gd. Although two transitions have
been observed in specific heat measurements at 
$T\approx 1$K and $T\approx 0.7$K \cite{gdtio-ramirez},
neutron scattering results only find the three sublattice
P state down to $T\approx 50$mK \cite{bramwell-private}.
We note that a $1$-${\bm k}$ structure is purposed experimentally 
for the P state, but a $4$-${\bm k}$ spin state 
is a possibility. In our RSMFT, a $4$-${\bm k}$ spin structure
corresponds to a unit cubic cell in which one of the tetrahedra
is fully disordered \cite{bramwell-scrystal}. 
It is unclear at this time whether it is a  
material (sample) specific effect or something else 
that causes the P state to freeze out in the neutron experiments 
as the material is cooled from 
$T\gtrsim 1$ K down to $50$mK \cite{gdtio-bramwell}.
Recent ESR results have also found peculiar behavior in
\Gd , where a large $[111]$ anisotropy develops spontaneously 
below $80$K \cite{gdtio-esr}.
Clearly, more experimental and theoretical work is needed to elucidate the
peculiar thermodynamic properties of \Gd.

In conclusion, our calculations above have 
identified another system, apart
from spin ice, where the long-range $1/r^3$ 
nature of dipolar interactions cause
a large degree of symmetry restoration on the 
${\bm q}-$dependent susceptibility.
However, this symmetry restoration is incomplete, 
and gives rise ultimately to a well defined ordering 
state upon cooling from the fully 
disordered paramagnetic state.


\begin{acknowledgments}
We would like to acknowledge Steve Bramwell, Ying-Jer Kao, 
Adrian del Maestro, Jean-Yves Delannoy, and Hamid Molavian 
for many useful discussions. This work is supported by NSERC 
of Canada, Research Corporation and the Province of Ontario.
\end{acknowledgments}

\bibliography{mft-gdtio.bib}

\begin{thebibliography}{20}
\bibitem[{num()}]{numerics}
\bibinfo{note}{Errors in the reported numerical values arise from finite size
  effects and round off errors associated with double precision floating point
  calculations and are less than $10^{-7}$}.
\bibitem[{bra()}]{bramwell-scrystal}
\bibinfo{note}{Neutron scattering on a single crystal would be needed to
  distinguish between a single-${\bm k}$ and a multiple-${\bm k}$ structure.}
\end{thebibliography}



\end{document}